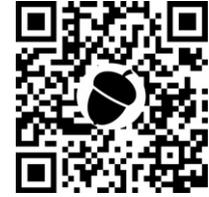

Open camera or QR reader and
scan code to access this article
and other resources online.

# A One-Dimensional Energy Balance Model Parameterization for the Formation of $CO_2$ Ice on the Surfaces of Eccentric Extrasolar Planets

Vidya Venkatesan,[1,*] Aomawa L. Shields,[1] Russell Deitrick,[2] Eric T. Wolf[3-5] and Andrew Rushby[6]

## Abstract

Eccentric planets may spend a significant portion of their orbits at large distances from their host stars, where low temperatures can cause atmospheric $CO_2$ to condense out onto the surface, similar to the polar ice caps on Mars. The radiative effects on the climates of these planets throughout their orbits would depend on the wavelength-dependent albedo of surface $CO_2$ ice that may accumulate at or near apoastron and vary according to the spectral energy distribution of the host star. To explore these possible effects, we incorporated a $CO_2$ ice-albedo parameterization into a one-dimensional energy balance climate model. With the inclusion of this parameterization, our simulations demonstrated that F-dwarf planets require 29% more orbit-averaged flux to thaw out of global water ice cover compared with simulations that solely use a traditional pure water ice-albedo parameterization. When no eccentricity is assumed, and host stars are varied, F-dwarf planets with higher bond albedos relative to their M-dwarf planet counterparts require 30% more orbit-averaged flux to exit a water snowball state. Additionally, the intense heat experienced at periastron aids eccentric planets in exiting a snowball state with a smaller increase in instellation compared with planets on circular orbits; this enables eccentric planets to exhibit warmer conditions along a broad range of instellation. This study emphasizes the significance of incorporating an albedo parameterization for the formation of $CO_2$ ice into climate models to accurately assess the habitability of eccentric planets, as we show that, even at moderate eccentricities, planets with Earth-like atmospheres can reach surface temperatures cold enough for the condensation of $CO_2$ onto their surfaces, as can planets receiving low amounts of instellation on circular orbits. Key Words: Extrasolar planets—Stars—Ice. Astrobiology 00, 000–000.

## 1. Introduction

The interaction of the host star spectral energy distribution (SED) with surface ice can significantly affect the planetary climate (Joshi and Haberle, 2012; Shields et al., 2013, Shields et al., 2014). While water ice has strong absorption at infrared (IR) wavelengths (3–150 μm), $CO_2$ ice is largely reflective at these wavelengths. These radiative differences are especially significant when considering the habitability of M-dwarf planets that emit strongly at near-IR wavelengths, where water ice absorbs strongly, which results in a warmer climate on M-dwarf planets than on planets orbiting stars with a higher ratio of visible/ultraviolet (UV) output (Shields et al., 2013, Shields et al., 2014; Shields et al., 2016). The radiative effects of $CO_2$ ice on the climate of eccentric planets orbiting stars of different spectral types may be significant and have not previously been explored.

---

[1]Department of Physics and Astronomy, University of California, Irvine, California, USA.
[2]School of Earth and Ocean Sciences, University of Victoria, Victoria, Canada.
[3]Laboratory for Atmospheric and Space Physics, University of Colorado Boulder, Boulder, Colorado, USA.
[4]Sellers Exoplanet Environment Collaboration (SEEC), NASA Goddard Space Flight Center, Greenbelt, Maryland, USA.
[5]Blue Marble Space Institute of Science, Seattle, Washington, USA.
[6]Department of Earth and Planetary Sciences, Birkbeck University of London, London, United Kingdom.
*NASA FINESST fellow.

Planets with eccentricities $e \geq 0.1$ comprise 20% of all discovered exoplanets with confirmed eccentricity[1] (Brady et al., 2018; Hugh et al., 2006; Kane et al., 2016; Kanodia et al., 2020; Korzennik et al., 2000; Naef et al., 2001; Schanche et al., 2022; Tamuz et al., 2008; Wittenmyer et al., 2017). Additionally, some rocky exoplanet candidates are also on eccentric orbits (Astudillo-Defru et al., 2020; Dreizler et al., 2020; Stock et al., 2020; Winters et al., 2022). At large orbital distances from their host stars, eccentric planets exhibit colder surface temperatures that may allow for the condensation of atmospheric species. Water vapor, having a high condensation temperature, would be the first species to condense at large distances. At even colder temperatures, "exotic" ices—ices other than water ice—could potentially condense on planetary surfaces. Carbon dioxide, nitrogen, ammonia, and hydrocarbon ices such as methane and ethane exist on planetary bodies within our own solar system, including Mars (Jones et al., 1979; Phillips et al., 2011), Triton (Cruikshank et al., 1993), and Pluto (Tegler et al., 2010) and its moon Charon (Brown and Calvin, 2000). On highly eccentric cold planets, bulk atmospheric constituents may condense onto the surface, resulting in the formation of large $CO_2$ ice deposits, similar to those observed on the polar regions of Mars (Smith et al., 2022). Water and $CO_2$ condensation on the surface can increase its reflectivity, which also increases planetary bond albedo and intensifies the water and $CO_2$ ice-albedo feedback. If the feedback is sufficiently strong, it could cause a planet to enter into a fully glaciated water snowball state, akin to snowball Earth (Hoffman et al., 1998; Kirschvink, 1992) or a fully glaciated $CO_2$ snowball state.

In some cases, eccentric rocky planets can support liquid water only for a portion of their orbit as they move to periastron due to the higher stellar flux they receive (Bolmont et al., 2016), while remaining in a frozen state at apoastron. However, the long-term climate stability of Earth-like planets depends on the orbit-averaged stellar flux received rather than their time spent within the habitable zone (HZ) (Kasting et al., 1993; Williams and Pollard, 2002). The HZ has an inner edge known as the moist greenhouse limit, where high atmospheric temperatures lead to an ineffective cold trap, which causes a moist stratosphere. At its extreme, the moist greenhouse limit of the HZ leads to a runaway greenhouse (RGH), resulting in uncontrolled warming due to the closure of the $H_2O$ window region. The outer edge is defined by the maximum $CO_2$ greenhouse limit, where the increased Rayleigh scattering in the multibar $CO_2$ atmosphere balances its greenhouse effect and prevents any further $CO_2$-induced warming of the planet (Kasting et al., 1993; Kopparapu et al., 2013, Kopparapu et al., 2014). However, Turbet et al. (2017) have shown that $CO_2$ surface ice condensation is more prominent at high solar constants than $CO_2$ maximum greenhouse limits. This study focuses on the RGH limit for planets. It utilizes the parameterization from Palubski et al. (2020) as it focuses on eccentric planets that may reach beyond the moist greenhouse limit and into the RGH regime at periastron.

Eccentric rocky planets, even outside the traditional HZ, may exhibit fractional habitability, and maintain liquid water on their surface during a portion of their orbits (Dressing et al., 2010; Linsenmeier et al., 2015; Palubski et al., 2020). In most studies, the mean flux approximation can be used to assess the overall climate (Bolmont et al., 2016), considering the average flux based on the orbital radius within the eccentric HZ, the region around a star where a planet on an eccentric orbit can maintain surface liquid water (Barnes et al., 2008). In certain cases, atmospheric condensation may occur at apoastron while intense stellar heating could potentially drive moist greenhouse conditions at periastron (Palubski et al., 2020). Using three-dimensional (3D) global climate model (GCM) studies, Bolmont et al. (2016) found that eccentric aquaplanets can freeze over at apoastron but recover their liquid water oceans at periastron. Research by Graham et al. (2022) indicates that at low installation (stellar flux from the host star), rocky planets within the circumstellar HZ can have condensed $CO_2$ oceans in addition to water oceans and that these $CO_2$ oceans can be present even in the presence of negative feedback due to silicate weathering. This might also be relevant for planets on eccentric orbits. Therefore, investigating the varying climates of eccentric planets throughout their orbits, from atmospheric condensation at apoastron to perhaps reaching RGH states imposed by the orbit-averaged flux, is crucial for enhancing our understanding of their long-term habitability.

Previous studies employing 1D energy balance models (EBMs) (Ramirez et al., 2020; Simonetti et al., 2024), 2D EBMs (Ramirez, 2024), and 3D General Circulation Models 9GCMs) (Forget et al., 2013; Soto et al., 2015; Urata and Toon, 2013; Wordsworth et al., 2013) have explored $CO_2$ condensation on Mars and select terrestrial exoplanets (Bonati and Ramirez, 2021; Turbet et al., 2017, Turbet et al., 2018). However, these studies neglected eccentric planets. Similarly, while EBM studies have examined the effects of water ice, salt, and land albedo feedback (Rushby et al., 2019, Rushby et al., 2020; Shields et al., 2013; Shields and Carns, 2018), the radiative impacts of $CO_2$ ice on the long-term climate stability of eccentric exoplanets have not been investigated.

In this study, we quantify the influence of the $H_2O$ and $CO_2$ ice-albedo feedbacks on eccentric and cold planets. The article is structured as follows: Section 2 explains the modifications made to incorporate a $CO_2$ ice-albedo parameterization into an EBM. In Section 3, we describe our multitiered approach to model the climate of cold and eccentric planets using our EBM, and we validate our model. In Section 4, we present the results obtained by varying eccentricity, host star SED, and $CO_2$ ice-grain size. Our findings and their implications are discussed in Section 5; conclusions follow in Section 6.

## 2. Methods

### 2.1. Model description

A 1D EBM from North and Coakley (1979) was used to simulate the potential climates of planets orbiting a range of stars at varying eccentricities. The EBM calculates the zonally averaged surface temperatures of the planet at each latitude and incorporates as input the albedos of ocean, land, and water ice surfaces, assuming the presence of an overlying atmosphere. We modified the EBM to also include an

---

albedo parameterization for the formation of $CO_2$ ice that may form on these planets' surfaces. The EBM equates the incoming shortwave radiation from the host star with the outgoing longwave radiation (OLR) from the planet using the following equation from North and Coakley (1979):

$$C(x)\frac{\partial T}{\partial t}(x,t) - D_0\Delta^2 T(x,t) + A + BT(x,t)$$
$$= QS(x,t)(1 - S(x,t)), \qquad (1)$$

In this equation, $x$ is the sine of the latitude, and $t$ is the time. The first term on the left-hand side of Eq.1 represents the time evolution of temperature, accounting for the thermal inertia or heat capacity per unit area of the surface of the planet. The second term captures the transport of heat across the planet, where $D_0$ serves as the diffusion parameter. The third and fourth terms jointly express the OLR emitted by the planet. The right-hand side of the equation describes the shortwave radiation absorbed by the planet. Here, $Q$ represents the globally averaged incident flux ($\sim 340$ Wm$^{-2}$), $S(x,t)$ denotes the latitude-dependent normalized flux received by the planet, and $A(x,t)$ signifies the albedo of the planet.

EBMs have been extensively used to study the ice-albedo feedback mechanism and its implications for Earth's paleoclimate (Budyko, 1969; Feldl and Merlis, 2021; Lindzen and Farrell, 1977). These models utilize the incoming shortwave radiation, OLR, and horizontal heat transport to determine the global mean temperature and mean iceline latitude of the planet (North and Coakley, 1979).

An Earth-like atmosphere was assumed, with a $CO_2$ concentration of 400 parts per million (ppm), typical of present-day levels. Additionally, we used a mid-latitude average water vapor concentration of 1% of the atmosphere at a constant surface pressure of 1 bar with a modern-day continental distribution. In this distribution, some portions of the latitude band are assigned the albedo of land, while others are assigned ocean albedos. The parameterization of A and B is determined by the linearization of the equations described by Spiegel et al. (2010). Additionally, the model has distinct heat capacities for land and ocean, and heat is allowed to flow between them. This is consistent with other studies, such as that by Wilhelm et al. (2022). The albedo values for land and ocean were calculated as weighted averages based on the fractions of land and ocean in each latitude band, taking into account the SED of the host star.

In our model, the term "broadband planetary albedo/bond albedo" refers specifically to the broadband (integrated across the full spectrum of wavelengths covered by the albedo values) albedo of the planet. The bond albedo, calculated using the Spectral Mapping Atmospheric Radiative Transfer Model (SMART) (Crisp et al., 1996) in Shields et al. (2013), includes the effects of atmospheric gases, clouds, and Rayleigh scattering. This definition therefore follows this approach to the traditional concept of "top-of-atmosphere (TOA) albedo" found in some radiative-convective model articles that follow this approach to include the contribution of clouds (Williams and Kasting, 1997). In our model, atmospheric effects have been incorporated into the water ice, ocean, and land surface bond albedos, as described by Shields et al. (2013). Hence, we refer to them as bond albedos. However, we used albedo values for pure $CO_2$ ice with no overlying atmosphere when the relevant $CO_2$ condensation temperature was reached on our simulated planets' surfaces; thus we refer to the albedo when $CO_2$ ice is present as "surface albedo." This approach has been followed in other studies such as those by Palubski et al. (2020), Deitrick et al. (2018), and Wilhelm et al. (2022). Once the condensation temperature for $CO_2$ is reached, it is presumed that the entire atmosphere, including water vapor, has condensed onto the surface, with $CO_2$ ice on the topmost layer of the surface (overlying ice or snow). While $N_2$ and $O_2$ would still be present in the atmosphere, their condensation points (70 and 90 K) are far lower than the condensation limit of $CO_2$ at its assumed partial pressure, and their contribution to the planetary bond albedo is minimal Ahrens et al. (2022); thus their radiative effects were neglected.

In a warm-start scenario, as the flux decreases, the temperatures on land decrease to 273 K, and in the ocean to 271 K, the freezing point of salt water on Earth, and the albedo values for land and ocean are increased to those for water ice. Thus, it is assumed that the land and ocean surfaces are becoming snow-covered and glaciated. For a partial pressure of $CO_2$ of 400 ppmv as used in this study, the condensation temperature is 131.06 K, at which point in the model $CO_2$ is presumed to condense, and the albedo values are then increased to those of $CO_2$ ice. For our cold-start cases, the model with the albedo values for $CO_2$ ice were initialized everywhere on the planet, since it is presumed that the atmospheric $CO_2$ has condensed on both land and ocean surfaces. As the instellation increases, the albedo values of $CO_2$ ice are decreased to that of water ice when the temperature reaches 271 K in the ocean and 273 K on land. As the planet thaws out of ice cover, the water ice-albedo values are decreased to those of ocean and land. In our model, the freezing point of water is depressed by salinity, which leads us to consider it as 271 K rather than the standard 273 K for the ocean. The obliquity was set to 0, while the instellation, eccentricity, host star SED, and $CO_2$ ice-grain size were varied to explore and isolate their climate effects. The complete set of model inputs is provided in Table 1.

TABLE 1. MODEL PARAMETERS

| Parameter | Type/value |
| --- | --- |
| Ice surface | $CO_2$, $H_2O$ |
| Stellar type | F2V, G2V, K2V, M3V |
| Rotation period | 24 h (F, G, K, M) |
| Dominant atmospheric species | $N_2$, $CO_2$, $H_2O$ |
| Atmospheric pressure | 1 bar |
| Amount of $CO_2$ | 400 ppmv |
| Stellar flux | 5–150%, varied by 5% |
| Obliquity | 0° |
| Argument of periastron | 102.07 |
| Eccentricity | 0.0, 0.5, 0.9 |
| Semi-major axis | 1 AU |
| Heat diffusion (D) | 0.44 Wm$^2$ K$^{-1}$ |
| Land configuration | modern Earth |
| $CO_2$ ice grain | 2, 20, 200, 2000 μm |
| Water ice grain | snow (S), 50% mixture of S + BM, blue marine (BM) |
| Heat capacity, land ($C_l$) | 0.45 Wm$^2$yr K$^{-1}$ |
| Heat capacity, ocean mixed layer ($C_w$) | 9.8 Wm$^2$yr K$^{-1}$ |
| Ocean mixed layer depth | 70 m |

## 2.2. Model inputs

The stellar spectra for F2V Star HD128167, the Sun, K2V Star HD22049, and M3V star AD Leo were retrieved from the Virtual Planet Laboratory's database (Chance and Kurucz, 2010; Reid et al., 1995; Segura et al., 2003; Segura et al., 2005). These stellar spectra are shown in Figure 1. The surface albedo files for water ice were obtained from Joshi and Haberle (2012). Snow corresponds to the condensation of water vapor from the atmosphere, blue marine ice forms from the freezing of liquid water and exhibits bubbles and cracks due to thermal stress (Warren et al., 2002), and the 50% mixture combines both snow and blue marine ice.

To examine the impact of ice-grain size on the climate stability of exoplanets, the $CO_2$ ice spectra of 2, 20, 200, and 2000 µm ice-grain sizes were obtained from Hansen (1997). These ice grains were grown under a partial pressure of $CO_2$ (35–40 mmHg) at a temperature of 150 K. Figure 2 showcases the surface spectra for snow, blue marine ice, the 50% mixture, and carbon dioxide ices, highlighting their different characteristics.

The bond albedo values for land, ocean, and water ice were obtained from Shields et al. (2013), derived from SMART which calculates the radiative physics between the planet's ocean, land, ice, and instellation based on the characteristics of the host star. The land and ocean grid cells exhibit a zenith angle dependence following Deitrick et al. (2018), which includes a Legendre polynomial to calculate the increased reflectance at high latitudes due to higher zenith angle. The bond albedo for water ice, land, and ocean includes the effect of Earth-like gases, clouds, and Rayleigh scattering. To simulate the $CO_2$ condensation on the planet's surface, the surface albedos were computed using the following equation:

$$\alpha = \frac{\int \alpha_\lambda \cdot F_{dn}(\lambda) d\lambda}{\int F_{dn}(\lambda) d\lambda},$$ (2)

Here, $\alpha$ is the bond albedo; $F_{dn}$ is the downwelling instellation reaching the surface, also known as "shortwave radiation," which is expressed as $\alpha_\lambda = F_{up}/F_{dn}$, where $F_{up}$ is the amount of flux that is scattered upwards and away from the planet's surface. The atmospheric effects were not included for $CO_2$ ices; hence, we refer to them as surface albedos.

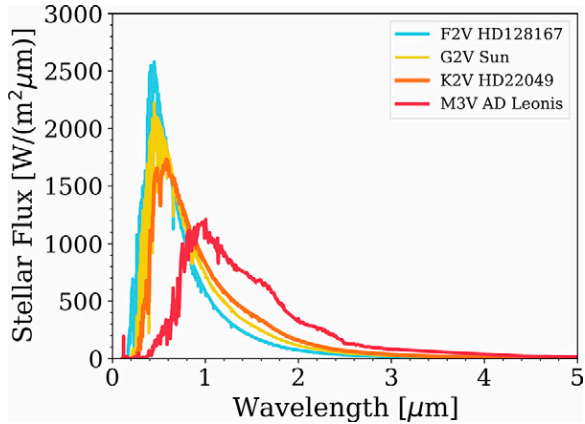

**FIG. 1.** SEDs for F-, G-, K-, and M-dwarf stars are normalized by integrated flux based on Wolf et al. (2017). SEDs, spectral energy distributions.

The input bond albedos and surface albedos are illustrated in Figure 3 and summarized in Table 2. These bond albedo values were utilized as input parameters for the EBM. We then used the EBM to examine the influence of eccentricity, host star spectral type, and ice-grain size on the climate sensitivity of planets at a range of eccentricities throughout the course of their orbits.

## 3. Introducing a Multitiered Approach to the EBM

The seasonal EBM computed the temperature and albedo for each latitude band at each time step throughout the year. The temporally resolved instellation representing the instantaneous flux at each time step was used for these calculations. To obtain annual mean temperatures and albedos, the time-varying values were averaged over the orbital period. To simulate cold planets, the seasonal EBM was modified to use a multi-layer approach. Starting with an Earth-like land and ocean configuration, and the OLR parameterization from Spiegel et al. (2010), the following traditional albedo parameterization was used for surface water ice when temperatures were below 273 K for land and below 271 K for ocean,

$$T = \begin{cases} \leq 271 \ K \ \& \ 273K, \text{water ice albedo} \\ >271 \ K \ \& \ 273K, \text{ocean and land albedo} \end{cases}$$ (3)

Once the daily temperatures dropped below the condensation temperature of $CO_2$ (131.06 K), the same OLR parameterization from Spiegel et al. (2009) was used, and the EBM was upgraded to include a parameterization for the formation of $CO_2$ ice, ice albedo

$$T = \begin{cases} \leq 131.06K, \ CO_2 \\ >131.06K, \text{water ice albedo} \end{cases}$$ (4)

Additionally, the RGH parameterization from Palubski et al. (2020) was incorporated, where the OLR was linearly parameterized based on surface temperatures,

$$OLR = \begin{cases} A + BT, T \leq 319K \\ 300 \ Wm^{-2}, T > 319K \end{cases}$$ (5)

where $A$ is 203.3 $Wm^{-2}$, $B$ is 2.08 $Wm^{-2}$, and $T$ is the zonally averaged surface temperature. For context, the OLR value for Earth is 239 ± 3 $Wm^{-2}$ (Stevens and Schwartz, 2012) compared with the RGH limit seen in the work of Palubski et al. (2020). Thus, these different parameterizations were applied to every latitude band.

Simulations were conducted in two scenarios: a "warm-start" case and a "cold-start" case. These cases are defined based on the orbit-averaged amount of instellation received by the planets from their host star. At higher eccentricities, the amount of instellation received by the planet averaged over the entire orbit increases by $(1 - e^2)^{-1/2}$, as demonstrated by past studies (Adams et al., 2019; Bolmont et al., 2016; Dressing et al., 2010; Laskar et al., 1993). We have incorporated this factor in our simulations and defined it as the orbit-averaged solar constant. For the cold-start case, the planet started in the globally ice-covered "$CO_2$ snowball" state with a global mean temperature below 131.06 K, and the instellation was increased by increments of 5% starting from 5% of the orbit-averaged solar constant. In these cases, it was assumed that the planets started out with $CO_2$ ice given their temperature regime and gradually warmed such that only water ice or liquid water was present, eventually transitioning into an RGH state when high values of insolation were reached.

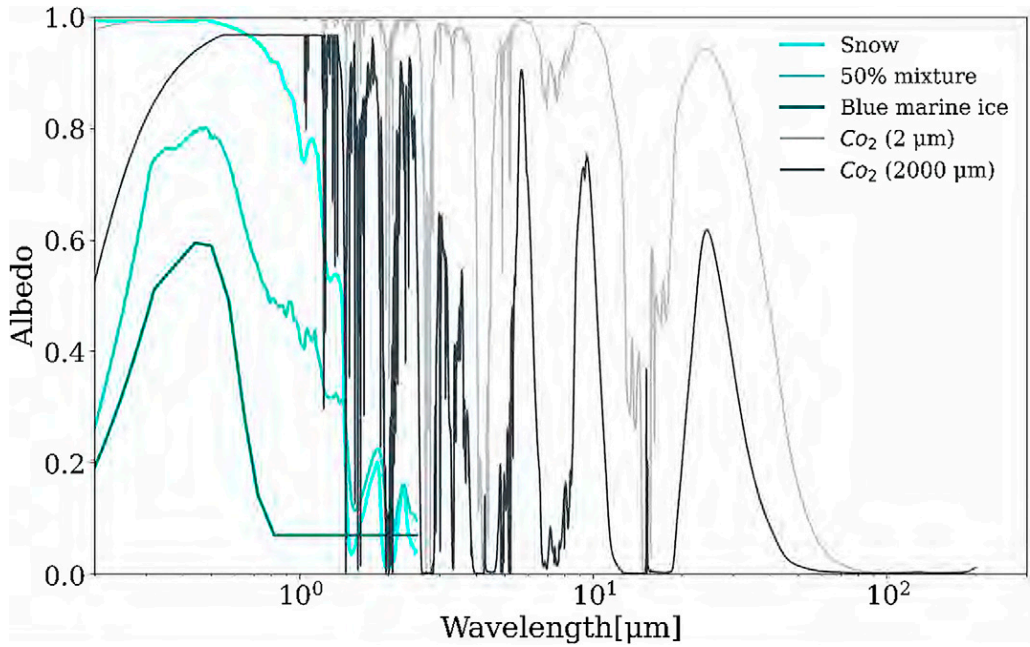

**FIG. 2.** $CO_2$ ice-albedo spectra were provided by Hansen (1997). Snow, blue marine ice, and the 50% mixture between the snow and blue marine ice spectra are from Joshi and Haberle (2012).

In contrast, for the warm-start cases, it was assumed that there was no ice on the planets at an instellation of 150% of the orbit-averaged solar constant. The instellation was then decreased in intervals of 5% until the planets reached fully glaciated, snowball states. As the instellation was further decreased, these planets transitioned from water ice snowball planets into $CO_2$ ice snowball planets. The RGH parameterization accounted for high temperatures, while water ice-albedo and $CO_2$ ice-albedo parameterizations were incorporated at lower temperatures that corresponded to the formation of water and $CO_2$ ice. The difference in the latitudinal extent of the ice line between the warm start and cold start is a measure of the planet's climate hysteresis (see, e.g., Broecker and Denton, 1989; Lenton et al., 2008). While our study focused on varying instellation while keeping the orbital period constant and Earth-like, research indicates that changes in the orbital period also influence climate hysteresis (Wilhelm et al., 2022). Figure 4 illustrates an example of this climate hysteresis curve and demonstrates the climate sensitivity to changes in instellation. A wider hysteresis loop suggests greater climate stability across a range of instellation. In contrast, a narrower loop indicates heightened sensitivity to changes in incoming stellar radiation, potentially leading to significant shifts in climate states.

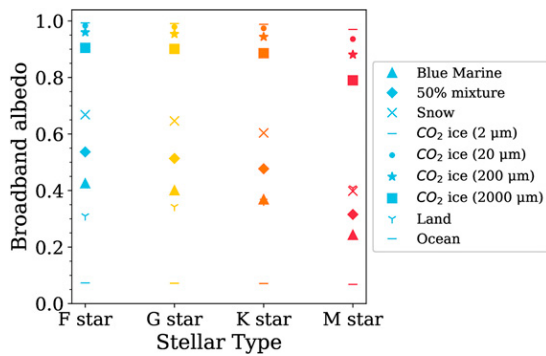

**FIG. 3.** Planetary bond albedo and surface albedo values are used as input to the EBM. Shown are bond albedos for snow, blue marine ice, and a 50% mixture of the two end-members, from Shields et al. (2013). Surface albedos for pure $CO_2$ ice of different ice-grain sizes are also shown. A fully condensed atmosphere is assumed at a surface temperature of 131.06 K when $CO_2$ forms, rendering these surface albedos equivalent to TOA albedos in that temperature regime, although $N_2$ and $O_2$ are still present in the atmosphere. EBM, energy balance model; TOA, top-of-atmosphere.

The model determined the annual percent surface coverage based on the resulting surface temperature output. Assuming a cold-start scenario, the model assigned the planet a surface albedo value characteristic of $CO_2$ ice since the surface temperature is below 131.06 K. As the instellation increased and the surface temperature rose above 131.06 K, the model switched to the albedo parameterization of water ice and the corresponding surface albedo for that surface type was used. As the planet thawed out of water ice, from 273 to 319 K, it was assumed that liquid water was present on the planet. A simple temperature-dependent albedo was used to simulate an ice sheet that started with an initial thickness of 2 m when the temperatures reached below −2°C. While the model tracks the thickness of water ice, it does not track the thickness of carbon dioxide ice. Given the lower condensation limit for $CO_2$ ice, any ice thickness output from this model, therefore, can be assumed to be a lower limit on what might actually accumulate on these planets' surfaces. Above 319 K,

| Host star | $CO_2$ ice | Surface albedo | $H_2O$ ice | Input albedo (Shields et al., 2013) |
|---|---|---|---|---|
| F2V | 2000 [μm] | 0.905 | Snow | 0.668 |
|  | 200 [μm] | 0.960 | 50% mixture | 0.536 |
|  | 20 [μm] | 0.983 | Blue marine | 0.425 |
|  | 2 [μm] | 0.993 | Land | 0.414 |
|  |  |  | Ocean | 0.329 |
| G2V | 2000 [μm] | 0.901 | Snow | 0.646 |
|  | 200 [μm] | 0.954 | 50% mixture | 0.514 |
|  | 20 [μm] | 0.979 | Blue marine | 0.401 |
|  | 2 [μm] | 0.993 | Land | 0.415 |
|  |  |  | Ocean | 0.319 |
| K2V | 2000 [μm] | 0.886 | Snow | 0.604 |
|  | 200 [μm] | 0.944 | 50% mixture | 0.447 |
|  | 20 [μm] | 0.974 | Blue marine | 0.369 |
|  | 2 [μm] | 0.988 | Blue marine | 0.401 |
|  | 2 [μm] | 0.991 | Blue marine | 0.401 |
|  | 2 [μm] | 0.993 | Land | 0.401 |
|  |  |  | Ocean | 0.302 |
| M3V | 2000 [μm] | 0.790 | Snow | 0.398 |
|  | 200 [μm] | 0.881 | 50% mixture | 0.315 |
|  | 20 [μm] | 0.936 | Blue marine | 0.243 |
|  | 2 [μm] | 0.970 | Blue marine | 0.401 |
|  | 2 [μm] | 0.991 | Blue marine | 0.401 |
|  | 2 [μm] | 0.993 | Land | 0.331 |
|  |  |  | Ocean | 0.233 |

the onset of an RGH state is presumed, following (Palubski et al., 2020) parameterization. Thus, at any given installation, the percent surface coverage for the different states can be determined on the planet.

### 3.1. Model validation

The model output was validated by reproducing the global mean temperature and ice lines for Earth and Mars. To validate the albedo effects of water ice, a 50% mixture of snow and blue marine ice on surfaces was assumed where temperatures fell below the freezing point of liquid water, as was done in previous studies (Rushby et al., 2019; Shields et al., 2013). The resulting ice line of 64° and a global mean temperature of 293 K were within 5 K of the results reported in previous studies (Rushby et al., 2019; Shields et al., 2013). The parameters used to validate Earth's OLR were taken from Spiegel et al. (2010), whose parameters are consistent with those of other studies (Deitrick et al., 2018; Palubski et al., 2020; Rushby et al., 2019; Shields et al., 2013; Shields and Carns, 2018; Wilhelm et al., 2022). A comparison between the TOA albedo and annual global mean temperature for our model versus Earth data is illustrated in Figure 5. Although there is a good agreement between our model and the Earth data, the TOA albedo appeared higher in our model. This difference is attributed to our consideration of a 50% mixture between snow and blue marine ice for surface ice, compared with sea ice used in their data. The transient increase observed in the Earth data corresponds to the presence of clouds. Our study used the same Spiegel OLR parameterization for Mars and Earth. However, we considered a 200 μm grain size of $CO_2$ ice and assumed that the planet received 43% of the orbit-averaged solar constant. Previous studies, such as that by Cross et al. (2020), have examined $CO_2$ ice samples on Mars with sizes ranging between ≥200 and 2000 μm. This study assumed that our simulated planets harbor a $CO_2$ ice-grain size of 200 μm. The model yielded a global mean temperature of 211 K, which aligns with current estimates of the surface temperature on Mars (Atri et al., 2022).

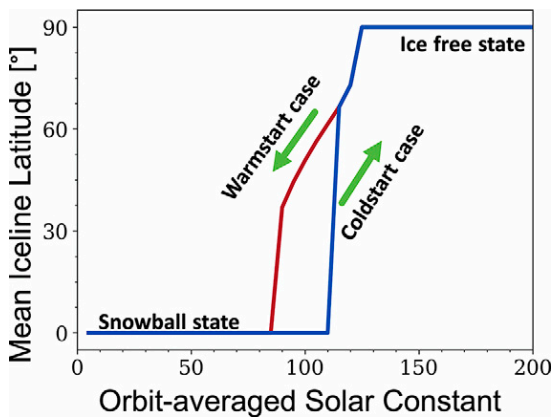

**FIG. 4.** A schematic diagram of a climate hysteresis curve. The two stable states—a globally glaciated snowball state and ice-free states—are shown in the figure. A warm-start case refers to a simulation that assumes the initial condition to be ice-free, whereas a cold-start case assumes the initial condition to be a snowball state.

We also followed the approach of Bonati and Ramirez (2021) and computed the northward heat flux for our model to validate our atmospheric heat transport parameter. The northward heat flux was calculated using the following equation:

$$F = -2\pi R^2 D cos^2(\lambda)\frac{dT}{dx}, \qquad (6)$$

where $F$ is the meridional heat transport flux, $R$ is Earth's radius (6371 Km), $\lambda$ is the latitude, $D$ is the heat coefficient and we used a fixed value of 0.44 W m$^{-2}$ K$^{-1}$, $T$ is the surface temperature, and $x = \sin\lambda$ represents the grid for each latitude band. The meridional heat flux as a function of obliquity at 0°, 23.3°, 45°, and 90° is depicted in Figure 6.

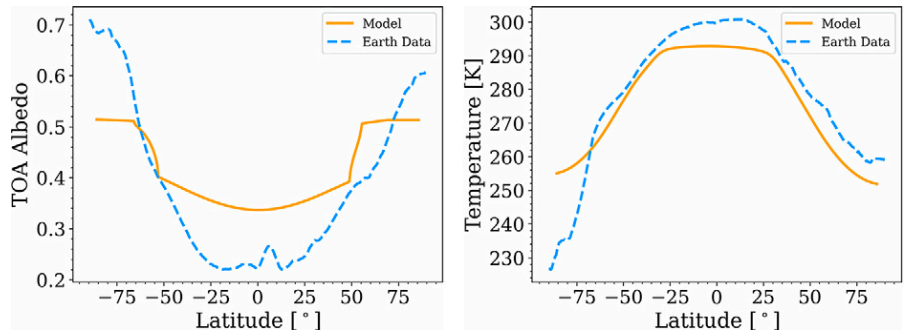

**FIG. 5.** A comparison between the results of our 1D EBM and observed data of Earth's top-of-atmosphere (TOA) albedo (left) and global mean surface temperature (right) is presented. The Earth's TOA albedo data are sourced from Energy balance and filled (CERES EBAF) (Loeb et al., 2018), while the annual surface temperature data were obtained from ERA5 reanalysis (Hersbach et al., 2020).

While our results are largely consistent with previous studies conducted by Williams and Pollard (2003), Linsenmeier et al. (2015), and Bonati and Ramirez (2021), a small anomaly at −40° latitude was observed, and a flat meridional heat transport pattern is present at an obliquity of 90°. These anomalies arise due to the modern-day land distribution assumed for the planet across latitudes. When the obliquity is 90°, the southern polar region is the substellar point so no gradient in temperature is observed at this point, resulting in a flux of 0 between −60° and −80° latitude. The −40° latitude coincides with the lowest fraction of land (5%), and the albedo is dominated by the ocean, which is darker than our land surface. This leads to higher temperatures at these points, resulting in a spike in the meridional heat transport. Although we tested different obliquities for model validation, all simulated planets in this study were assumed to have an obliquity of 0°. The validation of our Earth-based model, water ice parameterization, CO$_2$ ice-albedo parameterization, and zero obliquity studies of eccentric planets confirms the robustness of the model.

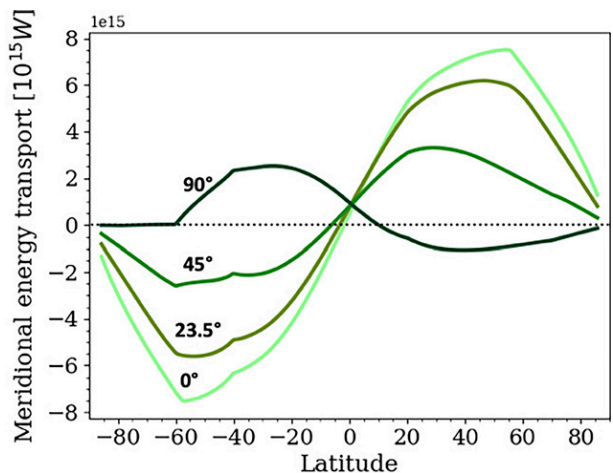

**FIG. 6.** The meridional energy transport as a function of latitude is shown for different obliquities. Different obliquities were tested for model validation; however, a 0° obliquity was considered for all simulations in our work.

## 4. Results

First, we compare simulations using our new CO$_2$ ice-albedo parameterization with those employing a traditional water ice-albedo parameterization. We determine the instellation threshold for CO$_2$ condensation on planets orbiting a range of different types of stars, and we examine how CO$_2$ and H$_2$O climate hysteresis is affected by eccentricity, host star SED, and different ice-grain sizes in exoplanets. Additionally, we investigate the instellation at which planets transition into RGH states.

Figure 7 presents the instellation at which CO$_2$ condensation occurs for warm-start planets orbiting F-, G-, K-, and M-dwarf stars. Warm-start planets with an eccentricity of 0.9 do not reach temperatures low enough to harbor CO$_2$ ice on their surface. M-dwarf planets with an eccentricity of 0.5 also fail to reach temperatures corresponding to the CO$_2$ condensation limit, even when the instellation is lowered to 5% of the orbit-averaged solar constant, the lowest limit we have used for the installation. In a warm-start scenario, F-, G-, K-, and M-dwarf planets on circular orbits require 5% of the orbit-averaged solar constant, equivalent to 68 W/m², for the formation of $CO_2$. With an increase in eccentricity from 0 to 0.5, the installation levels for $CO_2$ condensation are 5% of the orbit-averaged solar constant for F-, G-, and K-dwarf planets, respectively. This trend can be attributed to the longer wavelength emissions associated with the SEDs of cooler stars, which align with the strong absorption characteristics of $CO_2$ ice in those spectral regions.

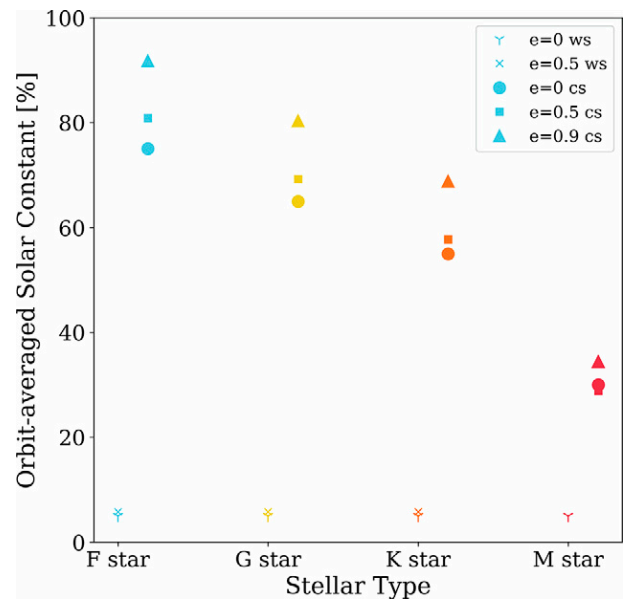

**FIG. 7.** The cross and Y markers quantify the instellation at which CO$_2$ condensation occurs for warm-start planets that get cold enough to harbor CO$_2$ ice on their surface. On the contrary, the instellation at which cold-start planets exit the CO$_2$ condensation limit is represented by circles, squares, and triangles.

In Figure 7, we also display the installation at which cold-start planets exit the $CO_2$ condensation limit. Cold-start planets begin with $CO_2$ ice and require a higher flux to melt the $CO_2$ ice on their surfaces due to increased albedos versus land and ocean surfaces. F-, G-, K-, and M-dwarf planets on circular orbits need 75%, 65%, 55%, and 30% of the orbit-averaged solar constant, respectively, to surpass the $CO_2$ condensation limit. Similarly, for F-, G-, K-, and M-dwarf planets with an eccentricity of 0.5, the required flux is 80.83%, 69.28%, 57.74%, and 28.87% to melt the $CO_2$ ice. However, highly eccentric planets ($e = 0.9$) can surpass the $CO_2$ condensation limit at higher flux levels. For instance, F-, G-, K-, and M-dwarf planets need 91.77%, 80.30%, 68.8%, and 34.4% of the orbit-averaged solar constant to melt $CO_2$ ice on their surfaces.

Figure 8 illustrates the latitudinal extent of water ice as a function of the orbit-averaged solar constant (1360 W/m²) for an F-dwarf planet and an M-dwarf planet, comparing the traditional water ice-albedo parameterization (teal) with the addition of a $CO_2$ ice-albedo parameterization (gray) in a cold-start scenario. Simulations assume a mixture of 50% snow and blue marine water ice, 200 μm $CO_2$ ice, and an orbital eccentricity of 0.5. With the inclusion of the $CO_2$ ice-albedo parameterization, climate simulations of M-dwarf planets require no difference in flux to exit global water ice cover than for simulations that use solely a water ice-albedo parameterization. However, F-dwarf planets are significantly more resistant to thawing and require an additional 29% increase in installation to exit global ice cover. The higher percentages of visible and near-UV flux emitted by F-dwarf stars interact with $CO_2$ and water ice on planetary surfaces; this results in higher bond albedos and a much larger requisite flux to thaw out of global ice cover than for M-dwarf planets. This demonstrates that including the $CO_2$ ice-albedo parameterization increases the installation required for thawing in F-dwarf planets compared with the water ice-albedo parameterization.

### 4.1. The effect of eccentricity

The hysteresis curves as a function of eccentricity for F-, G-, K-, and M-dwarf planets at $e = 0$, 0.5, and 0.9 are shown in Figure 9. These simulations included a 50% water ice mixture and 200 μm $CO_2$ ice-grain size. The width between warm- and cold-start conditions can be interpreted as a measure of climate sensitivity, where a narrower hysteresis indicates a higher climate sensitivity and a broader hysteresis implies a lower sensitivity to changes in installation. Initialized in warm-start conditions, a G-dwarf planet on a circular orbit required 85% of the orbit-averaged solar constant to transition into a snowball state, with the lowest ice line latitude migrating from 40.41° to 0°. However, G-dwarf planets with eccentricities of $e = 0.5$ and $e = 0.9$ did not appear to enter a fully glaciated condition unless the flux was reduced to 40% and 11.47% of the orbit-averaged solar constant, respectively. These planets also tended to have lower mean ice line latitudes. The periodic seasonal heating due to the high eccentricity orbit, coupled with the significant thermal inertia of the oceans, permits waterbelt states to remain stable.

In contrast, a fully glaciated Earth-like planet on a circular orbit required 110% of the orbit-averaged solar constant to exit a snowball state; the mean ice line latitude transitioned from 0° to 70.57°. Meanwhile, the planets with $e = 0.5$ and $e = 0.9$ required 69.28% and 80.30% of the orbit-averaged solar constant to thaw out of a globally ice-covered state. While the minimum ice line latitude changed from 0° to 10.91° for the $e = 0.5$ case, it switched from 0° to 18.4° in the case of the $e = 0.9$ simulations. Eccentric planets had a higher sensitivity to changes in installation, as seen in the data presented in Figure 9. Figure 10 displays the evolution of latitude distribution of heat flux received at periastron by a planet on a circular orbit compared with an eccentric planet ($e = 0.9$) such that highly eccentric planets received two orders of magnitude more flux, ~40,000 W/m², than a planet on a circular orbit, with implications for climate stability and long-term habitability, as discussed in the next section.

The climate sensitivity of eccentric planets was also correlated with the annual percent surface cover on the planet. The percentage of surface coverage by $CO_2$ ice, water ice, and liquid water for cold-start G-dwarf planets at different eccentricities is illustrated in Figure 11. When a G-dwarf planet is in a circular orbit, $CO_2$ ice is present up to 65% of the orbit-averaged solar constant. Between 65% and 100% installation, the planet has water ice on its surface, and in between 100% and 130% of the orbit-averaged solar constant, the planets are amenable to surface liquid water. However, above an installation of 130%, the planets enter a RGH state. If the eccentricity is increased to 0.5, $CO_2$ ice exists up to 68.5% of the orbit-averaged solar constant. Between 68.5% and 86.6% installation, the planet has temperatures amenable to the existence of water ice; and in between 86.6% and 121.24%, it can host liquid water, beyond which it enters an RGH state. Similarly, $CO_2$ ice is found on a planet with an eccentricity of 0.9 even at 80% of the orbit-averaged solar constant; between 80% and 137.6% of the orbit-averaged solar constant, it is possible for the planet to host either water ice or surface liquid water. However, beyond an orbit-averaged flux of 137.6%, the planet enters an RGH state. A similar trend emerges for F-, K-, and M-dwarf planets, where increasing eccentricity leads to a decrease in the annual ice fraction on planets, but the range of installations that exhibit surface conditions that support the presence of surface liquid water or water ice increases.

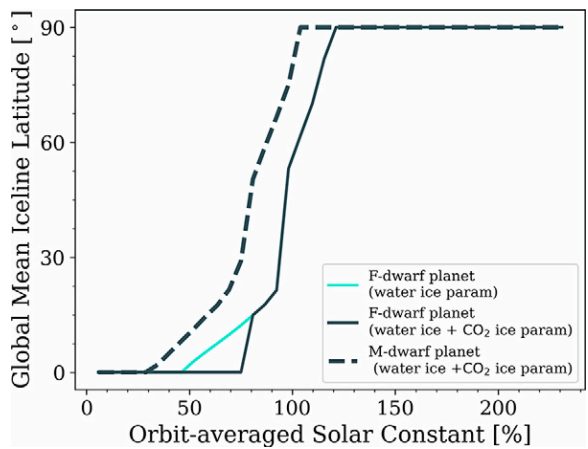

**FIG. 8.** Cold-start simulations are shown for an F-dwarf planet and an M-dwarf planet with an eccentricity of 0.5. The water ice-albedo parameterization is shown in teal, and the $CO_2$ ice-albedo parameterization is shown by dark gray.

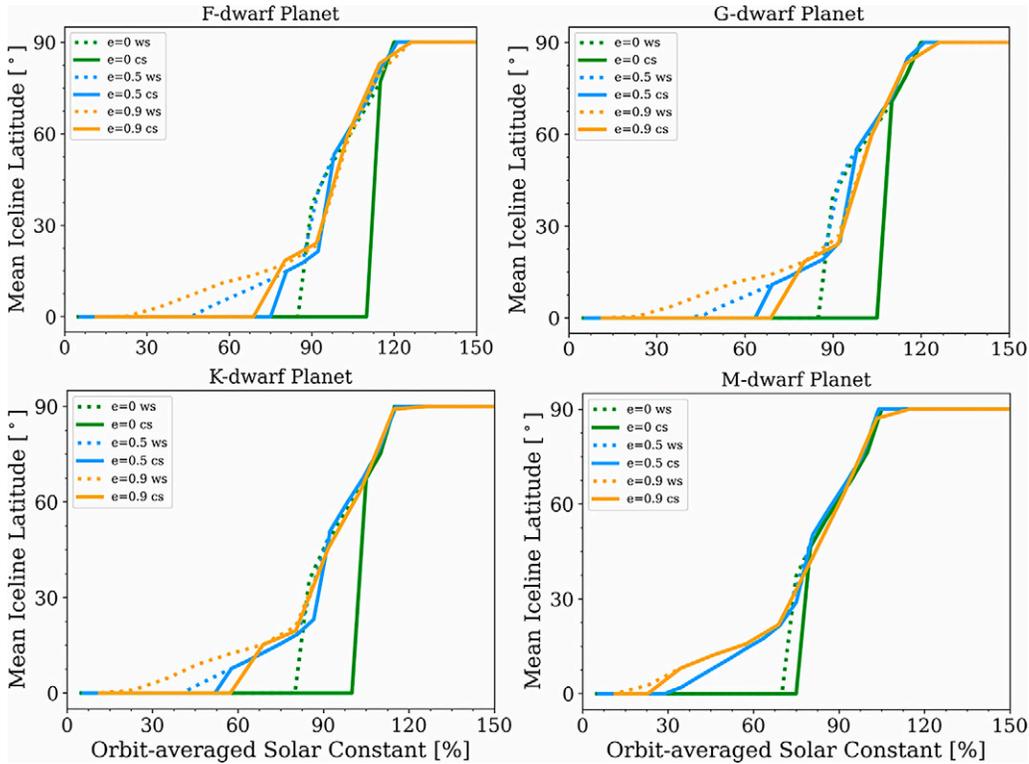

**FIG. 9.** The mean iceline latitude as a function of orbit-averaged solar constant for planets in warm-start and cold-start scenario orbiting F-, G-, K-, and M-dwarf host stars are shown. The warm-start cases are represented by dotted lines, and the cold-start cases are shown by solid lines. The green, blue, and orange lines correspond to eccentricities of 0, 0.5, and 0.9, respectively.

### 4.2. The effect of host star SED

The installation required for F-, G-, K-, and M-dwarf planets with $e = 0$, 0.5, and 0.9 to enter ice free or snowball conditions is shown in Figure 12. In the warm-start scenario, an F-dwarf and G-dwarf planet on a circular orbit requires 85% of the orbit-averaged solar constant to transition into a snowball state. The mean ice line latitude collapses to the equator from 36.4° to 39.7° for exoplanets hosted by F dwarfs and G dwarfs, respectively. K-dwarf and M-dwarf planets require 5% and 15% less flux than F-dwarf and G-dwarf planets to become globally glaciated.

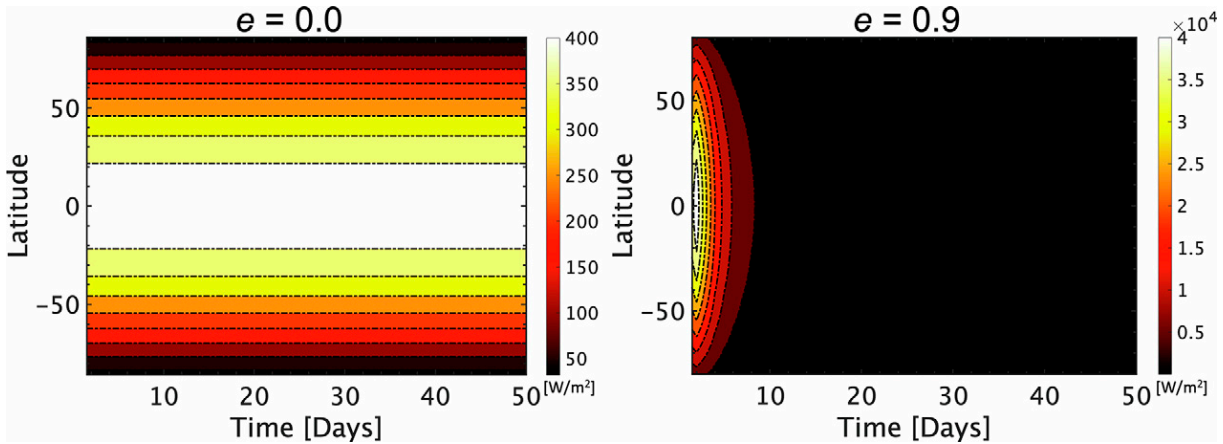

**FIG. 10.** The latitude distribution of received flux at periastron is shown for M-dwarf planets. A planet with an orbital eccentricity of 0.9 receives two orders of magnitude more flux than a planet on a circular orbit. While the simulations were run for 360 days, only the first 50 days of the orbit are displayed here to magnify the flux values for the $e = 0.9$ planet.

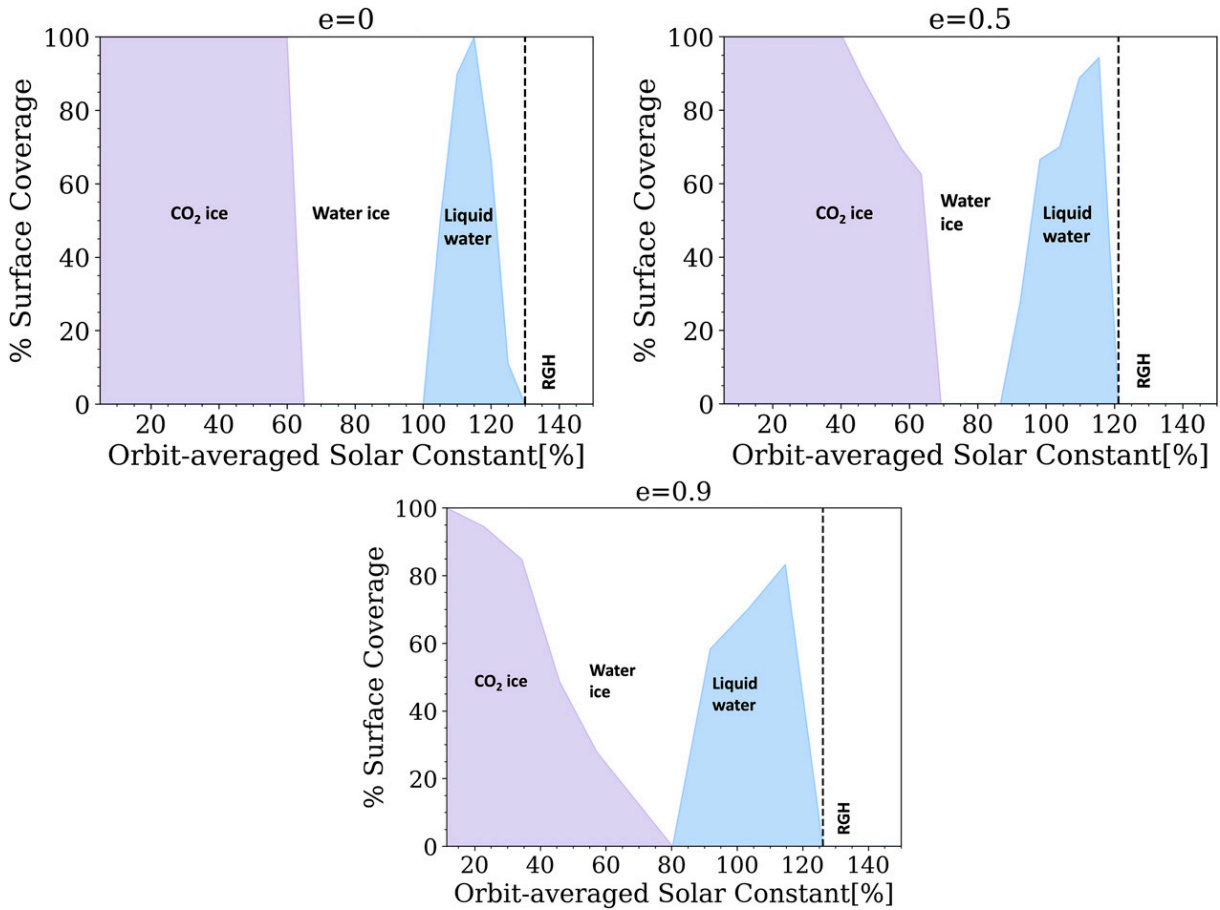

**FIG. 11.** Annual percent surface coverage for $CO_2$ ice, water ice, and liquid water are shown for G-dwarf planets with eccentricities $e = 0$, $e = 0.5$, and $e = 0.9$, respectively. Only the white region between purple and blue corresponds to water ice. The onset of the RGH effect is indicated by black dotted lines.

In the cold-start scenario, the effect of the $CO_2$ ice-albedo parameterization becomes especially prominent for planets orbiting brighter stars. An F-dwarf planet and a G-dwarf planet on a circular orbit require 115% and 110% of the orbit-averaged solar constant to transition from an water ice line of 0° to 77° and 70.57°, respectively. Likewise, K-dwarf and M-dwarf planets switched from a mean ice line latitude of 0° to 67.4° and 46.5° at 105% and 80% of the orbit-averaged solar constant, respectively, confirming a general trend that planets orbiting hotter, brighter stars are more resistant to thawing out of global ice cover, which applies to planets across all eccentricities. This trend is due to the larger percentages of visible and near-UV flux received from the host stars, which water ice (see, e.g., Shields et al., 2013, 2014) reflects strongly, and to the increased Rayleigh scattering on planets orbiting these stars (see, e.g., Kasting et al., 1993). The additional inclusion of $CO_2$ ice as we have done here, which is highly reflective at shorter wavelengths, also contributes to higher planetary bond albedos relative to planets orbiting cooler, less luminous stars.

### 4.3. Runaway greenhouse

Figure 13 displays the orbit-averaged flux at which the RGH transition occurs for F-, G-, K-, and M-dwarf planets with eccentricities of 0, 0.5, and 0.9, assuming an initial cold-start scenario. While F-, G-, and K-dwarf planets with an eccentricity of 0 require 125%, 120%, 120% of the orbit-averaged solar constant, respectively, to trigger an RGH, M-dwarf planets could enter an RGH state with 10% less installation (110% S) than G-dwarf and K-dwarf planets. A 50% increase in eccentricity does not significantly change the orbit-averaged flux required for planets to enter an RGH state. For example, the model suggests that F-, G-, K-, and M-dwarf planets require an installation of 121.24%, 121.24%, 121.24%, and 109.69% of the orbit-averaged solar constant to initiate the RGH feedback, respectively. F-, G-, K-, and M-dwarf planets on a highly eccentric orbit ($e = 0.9$) require considerably less installation, that is, 126.18%, 126.18%, 126.18%, and 114.7% of the orbit-averaged solar constant, respectively, to trigger an RGH. Overall, planets orbiting cooler stars required a smaller increase in installation to trigger an RGH feedback loop than planets with an eccentricity of 0 or 0.9, or planets orbiting brighter stars.

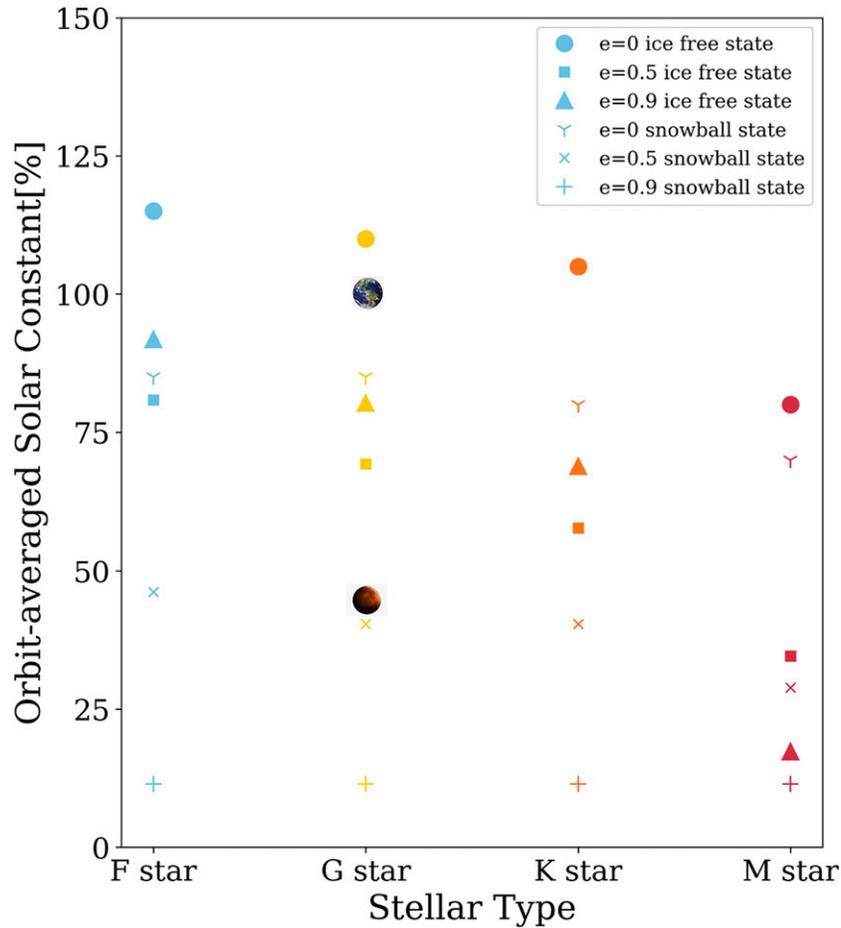

**FIG. 12.** The instellations required for warm-start planets to transition into a snowball state are shown in Ys, pluses, and crosses. The instellations required by cold-start planets to deglaciate at $e = 0$, 0.5, and 0.9 are displayed by circles, squares, and triangles. The orbit-averaged flux currently received by Earth and Mars, which is 100% and 43% of the orbit-averaged solar constant, respectively, is also shown for context.

### 4.4. Sensitivity to ice-grain size

$CO_2$ ices have been found on the poles of Mars in various forms, from fresh snow to thick ice deposits that have varying grain sizes (Hansen, 2005). The radiative properties of these different ice-grain sizes have been extensively studied in terrestrial laboratory settings (Chinnery et al., 2019; Hansen, 1997). In this study, simulations with ice-grain sizes of 2000, 200, 20, and 2 µm were performed to test the effect of varying ice-grain sizes on overall climate sensitivity. As the simulations attain an ice-free state for a cold-start case with a smaller instellation, this suggests larger ice grains aid in thawing, as also demonstrated in Figure 14. Figure 14 displays the effect of varying grain size on the climate hysteresis of an G-dwarf planet with an eccentricity of 0.5.

The warm-start case (not shown here) is not affected by the ice-grain size of $CO_2$ since it primarily depends on water ice-albedo parameterization, but the cold-start scenario, as shown in Figure 14, is strongly dependent on the $CO_2$ ice-albedo parameterization. Given simulations with cold-start conditions, planets with surface $CO_2$ ice with grain sizes of 2,

20, 200, and 2000 µm required 340.6%, 150.11%, 69.28%, and 46.19% of the orbit-averaged solar constant and transitioned from a mean iceline latitude of 0° to 90°, 90°, 10.91°, 1.06°, respectively, to enter an ice free state. This suggests a strong connection between climate sensitivity and ice-grain size. The trend of larger ice-grain sizes requiring less instellation to exit a globally glaciated snowball state persists across planets orbiting F-, G-, K-, and M-dwarf stars, and across all eccentricities.

### 5. Discussion

Our simulation study demonstrated that as eccentric planets experience cold enough temperatures for the condensation of atmospheric $CO_2$ and the formation of both $CO_2$ ice and water ice on their surfaces, such planets experience a significant climatic impact when a $CO_2$ ice-albedo parameterization is included in a 1D energy balance climate model. This finding underscores the importance of incorporating albedo parameterizations for the formation of surface $CO_2$ ice into climate simulations for planets with non-zero eccentricities. Such parameterizations in modeling can yield more accurate insights into the potential climates of these diverse worlds. Moreover, this parameterization can extend to the analysis of Mars, cold exoplanets, and planets positioned at substantial distances from their host stars, where conditions permit $CO_2$ surface ice formation.

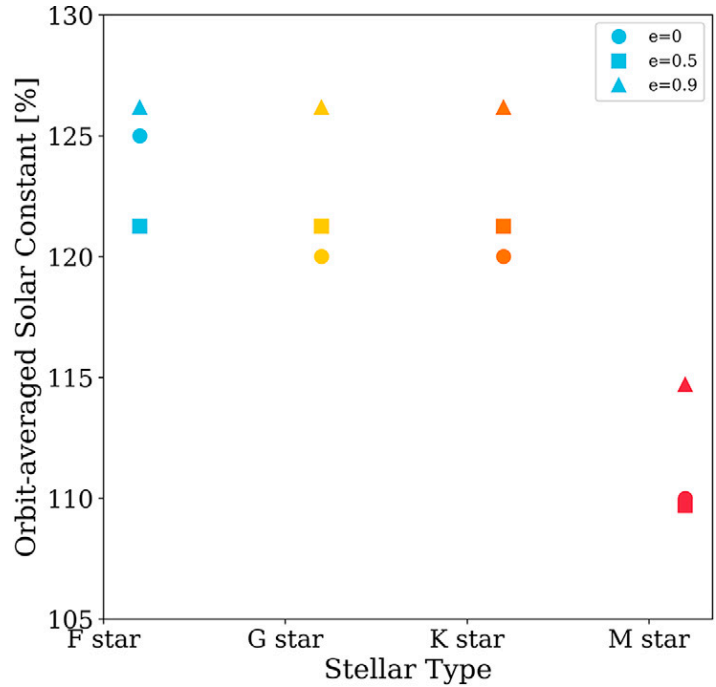

**FIG. 13.** The instellations at which planets enter an RGH state for cold-start planets at different eccentricities are shown in circles, squares, and triangles. Planets orbiting M-dwarf stars require less instellation to enter an RGH state than planets orbiting F-, G-, or K-stars.

Our simulations that included a $CO_2$ ice-albedo parameterization reveal an amplified climate hysteresis for all planets compared with the simulations that only used a water ice-albedo parameterization. This amplification is attributed to the higher bond albedo of $CO_2$ ice compared with that of water ice. Studies on Mars typically utilize an albedo value of approximately 0.6 for $CO_2$ ice (Turbet et al., 2018). However, it is important to note that the observed $CO_2$ ice albedo values on Mars can vary significantly due to dust presence, with dust notably decreasing the albedo of pure $CO_2$ ice (Kieffer, 1990; Kieffer et al., 2000). Consequently, the values reported in the literature are often lower than estimates for pure $CO_2$ ice. A mixture of dust and ice would lead to a lower albedo and a narrower climate hysteresis. However, in this study, it is presumed that all atmospheric constituents that would have an appreciable effect on the TOA albedo have condensed out onto the surface if the condensation point of $CO_2$ is reached. Our albedo values and spectra align with studies by Singh and Flanner (2016), where the authors utilized an albedo model for pure dust-free carbon dioxide snow albedo and applied it to Mars.

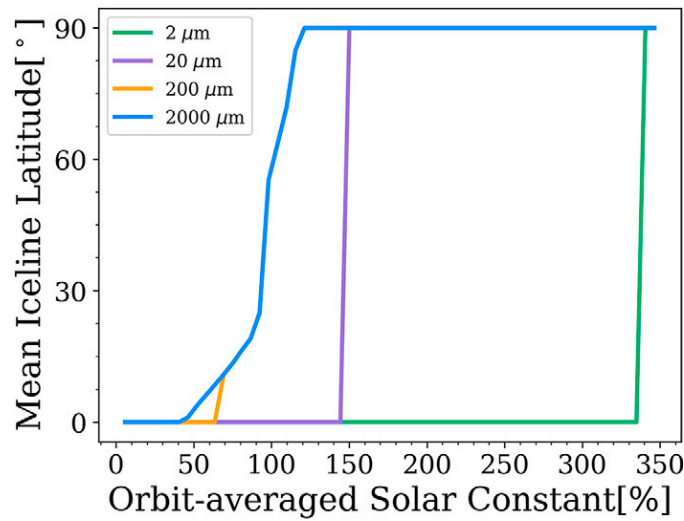

**FIG. 14.** The flux required for cold-start G-dwarf planets with an eccentricity of 0.5 and with ice-grain sizes of 2000, 200, 20, and 2 μm, respectively, to transition out of global ice cover is shown.

Kadoya and Tajika (2019) argue that the value of the $CO_2$ ice albedo should not affect the onset of snowball states on planets because it depends primarily on water ice and not on $CO_2$ ice, and we found that this is true for warm-start planets. However, our $CO_2$ ice-albedo values affect the amount of instellation required to exit global ice cover on cold-start planets, when one assumes $CO_2$ is already condensed across the surface. For example, Figure 14 illustrates the significant change in instellation required for G-dwarf planets in cold-start scenarios to transition out of a snowball state. Although the difference in values is relatively small (0.79 and 0.97 for 2000 and 2 μm endmembers, respectively) the increase in instellation required to exit global $CO_2$ ice cover is significant (250%). The higher stellar flux threshold requirements as ice-grain sizes reduce is a direct consequence of the intense ice-albedo feedback. We have modeled the planets assuming that the atmospheric $CO_2$ does not change because it is assumed that there is no carbon-silicate cycle present on these planets, which adjusts the silicate weathering rate with temperature (Walker et al., 1981). Including this effect will reduce the greenhouse effect when $CO_2$ condenses, leading to even colder climates and even greater instellations needed to exit a snowball state. Conversely, we have only studied one (low) value of $CO_2$. A 1 bar $CO_2$ atmosphere, for instance, would have a condensation temperature of 194 K and also feature a much stronger greenhouse effect, both of which would alter our results. These are an interesting topic for a future study.

The increased radiation received by eccentric planets at periastron promotes the melting of ice and impedes further ice growth on a planet's surface. As demonstrated in Figure 11, a planet with an eccentricity of 0.9 receives two orders of magnitude more flux at periastron than a planet on a circular orbit, as described by $F \propto (1 - e^2)^{-1/2}$. On these planets, $CO_2$ ice only forms on the surface once the instellation has been lowered to 5% of the orbit-averaged solar constant, which is our lower limit for instellation. In contrast, planets with $e = 0.5$ form $CO_2$ ice with an instellation of 20% of the modern solar constant.

As eccentricity increases, annual ice fraction for $CO_2$ ice decreases, as shown in Figure 11, but the range of instellations that exhibit surface conditions that support the presence of surface liquid water or water ice increases, which is in accordance with Palubski et al. (2020). Thus, highly eccentric planets may exhibit warmer surface conditions along a broader range of instellations. This is because the intense heat received at periastron aids these planets in melting the ice and maintains an overall warm climate. This also suggests that planets with extreme eccentricity are less likely to enter a snowball state due to a weaker ice-albedo feedback. This results in a narrower hysteresis curve, which allows for deglaciation with a smaller increase in instellation. Furthermore, it implies that eccentric planets are less prone to being trapped in a snowball state compared with planets on a circular orbit.

An RGH state is observed for all planets at a comparable orbit-averaged instellation of around 120% of the orbit-averaged solar constant. M-dwarf planets receive about two orders of magnitude higher orbit-averaged flux compared with planets with eccentricities of 0.9. Still, they enter an RGH state at about similar instellation as their counterparts.

This propensity is attributed to the strong absorption of $CO_2$ in the IR, where M dwarfs emit strongly. Notably, eccentric M-dwarf planets have also exhibited habitable surface conditions for a larger portion of their orbits compared with planets orbiting stars of different spectral types (Palubski et al., 2020). This suggests that eccentric M-dwarf planets might experience longer periods of fractional habitability during their orbits and maintain favorable conditions for life.

The $CO_2$ ice-albedo parameterization increases the amount of stellar flux required for planets orbiting F-, G-, and K-stars to exit a snowball state, compared with those orbiting M-dwarf stars. This distinction arises because M dwarfs exhibit strong IR emission, which corresponds to IR absorption and the subsequent melting of $CO_2$ ice on the surface. In contrast, F- and G-dwarf stars have a higher UV output, which causes $CO_2$ ice to be more reflective, thereby reducing surface temperatures and increasing the ice growth on these planets. Turbet et al. (2017) suggested that $CO_2$ condensation can impede the carbonate-silicate cycle on planets orbiting Sun-like stars, depleting atmospheric $CO_2$ and raising the likelihood of the planet entering a permanent snowball state. Our findings add an additional complication, as $CO_2$ ice increases the amount of flux required to thaw out of global ice cover for F- and G-dwarf planets more than for K- and M-dwarf planets.

While the significance of the $CO_2$ ice-albedo parameterization is evident, an associated challenge in incorporating it into every climate model lies in its strong dependence on grain size. In this study, it is demonstrated that planets with larger $CO_2$ ice grains may be able to deglaciate from a snowball state at lower levels of equivalent stellar flux due to their reduced efficiency of light scattering, which results in a lower surface albedo (Hansen, 2005). This behavior weakens the ice-albedo feedback, a critical factor in the formation and maintenance of a snowball state. Consequently, planets with larger $CO_2$ ice grains may undergo thawing and transition out of a globally glaciated state with a relatively small increase in instellation.

The analysis also reveals that the warm-start planets remain unaffected by changes in the grain size of the $CO_2$ ice due to the higher albedo of water ice on these planets. As discussed, F-, G-, K-, and M-dwarf planets with can eccentricity of $e = 0.5$ do not form $CO_2$ ice. This implies that the average instellation effect $((1 - e^2)^{-1/2})$ is more significant than the seasonal apoastron effect $\left( \frac{F_{max}}{F_{min} = \frac{(1+e)^2}{(1-e)^2}} \right)$ for highly eccentric planets.

The ability of cold-start planets to escape a global ice cover is heavily influenced by the albedo values of $CO_2$ ice. For example, a cold-start G-dwarf planet with $e = 0.5$ consisting of a 2000 μm $CO_2$ ice-grain size requires a flux equivalent to 40% of the orbit-averaged solar constant to emerge from a snowball state, while a similar planet composed of 2-μm grains requires 295% of the orbit-averaged solar constant.

In this study, it was assumed that $CO_2$ ice forms when the temperature drops below its condensation point and accumulates on top of water ice. While this approximation is suitable for an EBM study, it is important to note that $CO_2$ ice deposits may be less stable than water ice due to the former's higher density, so it could potentially remain beneath the water ice layer for up to $10^4$ years (Turbet et al., 2018). Given that we focus on simulations over an annual cycle (360-day orbit), our assumption is reasonable. This longevity could slow or even prevent deglaciation on these planets. Additionally, our research made the assumption of an Earth-like rotation and orbital period for all planets. Synchronously rotating warm-start planets are likely to undergo intense cloud formation on the day side, which would raise the albedo significantly (Yang et al., 2013). Including this effect for warm-start planets in our study would increase the amount of flux required for M-dwarf planets to exit a globally covered snowball planet. Note that these results only test a single value of $CO_2$, 400 ppm, which provides very little greenhouse warming and also results in a very low condensation temperature of $CO_2$. Outside of this study, we also tested simulations (not included in the article) where a 1 bar $CO_2$ atmosphere was assumed, and we found that it would lead to a broader climate hysteresis for all planets since the amount of installation required to exit a globally glaciated planet would increase. We employed spectra of pure $CO_2$ ice for our work. While surface ice may include constituents such as continental and volcanic dust (Abbot and Pierrehumbert, 2010), the results that use these idealized spectra provide a reasonable approximation. Last, we did not use a dynamic evolution of OLR in our study. When $CO_2$ begins to condense onto the surface, the amount of $CO_2$ in the atmosphere will decrease with time, leading to changes in the OLR coefficients as the greenhouse effect decreases. This dynamic evolution in OLR would be beneficial to explore in future work.

It is crucial to incorporate our ice-albedo parameterization for the formation of surface $CO_2$ ice into simulations of cold planets, especially given the recent James Webb Space Telescope (JWST) results which suggest the possible absence of atmospheres for inner M-dwarf planets like Trappist 1b and c (Greene et al., 2023; Zieba et al., 2023). These cold M-dwarf planets are in the Outer edge of habitable zone (OHZ) and are intriguing targets for characterization. Furthermore, characterizing planetary bond albedo could lead to the differences in the phase curves (Kane and Gelino, 2010) on these planets, which can be amenable to reflected light observations via the Habitable Worlds Observatory (Engineering National Academies of Sciences and Medicine, 2021) and Large Interferometer for Exoplanets (Defrère et al., 2018).

Despite the extreme cold temperatures experienced by our simulated planets, such conditions might not be detrimental, particularly during the transition from apoastron to periastron. Notably, episodes of global glaciation, like the Neoproterozoic snowball Earth events occurring between 750 and 635 million years ago, have been associated with the emergence of multicellular life on Earth (Planavsky et al., 2010). These environments can harbor a range of purple bacteria that are found in cold environments on Earth and could potentially survive on a snowball Earth orbiting a cooler star (Fonseca Coelho et al., 2024).

Thus, eccentric planets, planets at the OHZ and large distances from their host stars could host these exotic ice surfaces and provide exciting prospects for future observation, characterization, and habitability.

## 6. Conclusions

In this work, we used a 1D EBM with a novel ice-albedo parameterization for surface $CO_2$ ice formation to explore its radiative effects on the climates of Earth-like planets orbiting F-, G-, K-, and M-dwarf stars. Our findings highlight the need for more accurate albedo parameterizations for the formation of exotic ices into climate model simulations of eccentric planets, which can reach temperatures conducive to the condensation of atmospheric $CO_2$, leading to $CO_2$ ice forming on their surfaces. Our model simulations indicate that incorporating parameterization for the formation of $CO_2$ ice significantly impacts planetary climate. These climate effects are sensitive to eccentricity, host star spectral type, and ice-grain size. We demonstrated that ice-covered planets require higher installation levels to thaw from global $CO_2$ ice cover when incorporating the $CO_2$ ice-albedo parameterization, particularly for F-, G-, and K-dwarf planets compared with M-dwarf planets. The inclusion of radiative $CO_2$ ice-albedo effects in our simulations amplifies the extended climate hysteresis trend of a larger increase in installation to deglaciate for planets around F-dwarf stars in contrast to M-dwarf planets. Eccentric planets thaw out of global ice cover with a smaller relative increase in flux than planets on circular orbits because the intense flux (2 orders of magnitude higher) at periastron renders eccentric planets less susceptible to a snowball state. A larger ice-grain size of 2000 μm reduces the installation required to exit global ice-covered conditions, due to reduced light scattering and lower overall planetary albedo. Our work highlights the importance of integrating a $CO_2$ ice-albedo parameterization into climate simulations of diverse planetary environments, such as those found on Mars, cold and eccentric exoplanets, as well as those at large orbital distances from their host stars. Future investigations, using EBMs and GCMs, are anticipated to explore the climates of these cold worlds, especially those at the OHZ. This offers exciting observational prospects for current and upcoming telescopes.

### Acknowledgments

Cecilia Bitz originally coded the EBM. Special thanks to current Shields Center for Exoplanet Climate and Interdisciplinary Education team member Ana Lobo and former members Igor Palubski and Nick Duong. V.V. is grateful to Aaron Donohoe from the University of Washington Seattle for providing the TOA albedo and annual surface temperature data for the Earth. A.L.S. is grateful to the Virtual Planetary Laboratory at the University of Washington for fostering long-term collaborations that resulted in this article. The authors would like to thank the two anonymous reviewers at *Astrobiology* for their invaluable feedback that improved this article.

### Data Availability Statement

All data products and spectra used in this study are available via the Exoclimates Database maintained by the Shields Center for Exoplanet Climate and Interdisciplinary Education (SCECIE) at UC Irvine.

### Author Disclosure Statement

No competing financial interests exist.

## Funding Information

This material is based upon work supported by the NASA FINESST fellowship Grant no. 80NSSC21K1852 and NSF Award no. 1753373.

Address correspondence to:
*Vidya Venkatesan*
*Department of Physics and Astronomy*
*University of California*
*4129 Frederick Reines Hall*
*Irvine*
*CA 92697-4575*
*USA*

*E-mail:* vidyav1@uci.edu

**Abbreviations Used**

EBM = Energy Balance model
EBAF = Energy balanced and filled
GCMs = General Circulation models
HZ = Habitable Zone
IHZ = Inner Habaitable Zone
JWST = James Webb Space Telescope
OHZ = Outer Habitable Zone
OLR = Outgoing Longwave Radiation
RGH = Runaway Greenhouse
SMART = Spectral Mapping Atmospheric Radiative Transfer Model
TOA = top-of-atmosphere